\documentclass[aps,pre,floats,floatfix,twocolumn]{revtex4}

\usepackage{ifthen}
\usepackage{ifpdf}
\usepackage{color}

\ifpdf
\usepackage{graphicx}
\usepackage{epstopdf}
\else
\usepackage{graphicx}
\usepackage{epsfig}
\fi
\graphicspath{{./Figs/}{./}}

\usepackage{latexsym}
\usepackage{amsmath}
\usepackage{amssymb}
\usepackage{bm}
\usepackage{wasysym}

\usepackage{hyperref}

\newcommand{\eexp}{\mbox{e}^}

\newcommand{\amatrix}[1]{\begin{matrix} #1 \end{matrix}}

\newcommand{\mylabel}[1]{\label{#1}} 
\newcommand{\beq}{\begin{eqnarray}}
\newcommand{\eeq}{\end{eqnarray}} 
\newcommand{\be}[1]{\begin{eqnarray}\ifthenelse{#1=-1}
{\nonumber}{\ifthenelse{#1=0}{}{\mylabel{e#1}}}}
\newcommand{\ee}{\end{eqnarray}} 

\newcommand{\Eq}[1]{\textcolor{blue}{Eq.\!\!~(\ref{#1})}} 
\newcommand{\Fig}[1]{\textcolor{blue}{Fig.}\!\!~\ref{#1}} 
\newcommand{\hide}[1]{}
\newcommand{\rmrk}[1]{{#1}}    %{\textcolor{red}{#1}}
\renewcommand{\cite}[1]{\textcolor{blue}{[\onlinecite{#1}}]} %{[\onlinecite{#1}]} 

\begin{document}

\title{Non-equilibrium steady state and induced currents of a mesoscopically-glassy system:\\
interplay of resistor-network theory and Sinai physics}

\author{Daniel Hurowitz$^1$, Saar Rahav$^2$, Doron Cohen$^1$}

\affiliation{
\mbox{$^1$Department of Physics, Ben-Gurion University of the Negev, Beer-Sheva, Israel} \\
\mbox{$^1$Schulich Faculty of Chemistry, Technion - Israel Institute of Technology, Haifa 32000, Israel}
}

\begin{abstract}
We introduce an explicit solution for the non-equilibrium steady state (NESS) 
of a ring that is coupled to a thermal bath, and is driven by an external 
hot source with log-wide distribution of couplings. 
Having time scales that stretch over several decades is similar to glassy systems.
Consequently there is a wide range of driving intensities where the NESS 
is like that of a random walker in a biased Brownian landscape. 
We investigate the resulting statistics of the induced current~$I$. 
For a single ring we discuss how the sign of $I$ fluctuates as the intensity 
of the driving is increased, while for an ensemble of rings we highlight 
the fingerprints of Sinai physics on the distribution of the absolute value of $I$.   
\end{abstract}

\maketitle

%%%%%%%%%%%%%%%%%%%%%%%%%%%%%%%%%%%%%%%%%%%%%%%%%%%%%%%%%%%%%%%%%%%%%%%%%

\section{Introduction}

The transport in a chain due to random non-symmetric transition probabilities
is a fundamental problem in statistical mechanics \cite{derrida,d1,d2,d3,d4,sinai,sinai2}.
This type of dynamics is of great relevance for surface diffusion \cite{surface}, 
thermal ratchets~\cite{brownian1,brownian2,brownian3,ratchets}  
and was used to model diverse biological systems, 
such as molecular motors, enzymes, and unidirectional motion 
of proteins along filaments \cite{motors1,brownian4,motors2,motors3}.
Of particular interest are applications that concern the conduction 
of DNA segments \cite{DNA1,DNA3}, and thin glassy electrolytes under high voltages 
\cite{Wichmann,Roling1,Roling2,Roling3,Nitzan}.

Mathematically one can visualize the dynamics 
as a {\em a random-walk in a random environment}: 
a particle that makes incoherent jumps between ``sites" of a network.
In an unbounded quasi-one-dimensional network we might have either diffusion 
or sub-diffusive Sinai spreading \cite{sinai}, 
depending on whether the transitions rates 
form a symmetric matrix or not. In contrast, when the system is bounded 
(and without disjoint components) 
it eventually reaches a well-defined steady state. 
This would be an equilibrium {\em canonical} (Boltzmann) state if the transition rates 
were detailed-balanced, else it is termed non-equilibrium steady state (NESS).

Considering the NESS of a mesoscopically glassy system, 
{\em our working hypothesis is that glassiness might lead to a novel NESS
with fingerprints of Sinai physics.}
By ``glassiness" we mean that the rates that are induced by a bath, 
or by an external source, have a log-wide distribution,
hence many time scales are involved \cite{Rit1} as in spin-glass models \cite{Rit2}.  
Having a log-wide distribution of time scales is typical 
for hopping in a random energy landscape, 
where the rates depend exponentially on the barrier heights.
It also arises in driven quasi-integrable systems, 
where due to approximate selection-rules there is a ``sparse" fraction 
of large coupling-elements, while the majority become very small \cite{slk}.

The emergence of Sinai physics in a system that is described 
by a rate equation with asymmetric transition probabilities 
is not self-evident \cite{Rit3}.
An experimental observation of Sinai diffusion regarding the unzipping 
transition of DNA molecules has been reported \cite{Nelson},
and other applications have been considered \cite{domainwalls,sandpiles}.
The non-linear current dependence of a mesoscopic rings has 
been theoretically studied in the past \cite{Wichmann,Nitzan}, 
with references to experiments \cite{Roling1,Roling2,Roling3}, 
{\em but the statistical aspects, and the possible relevance 
of Sinai physics, have not been considered.}
In previous publications, we have pointed out that due to ``glassiness" 
Sinai physics becomes a relevant ingredient in the analysis of energy 
absorption \cite{kbb} and transport \cite{ner} in such a ring system.

%%% This work

\begin{figure}
\centering
\includegraphics[width=0.5\hsize]{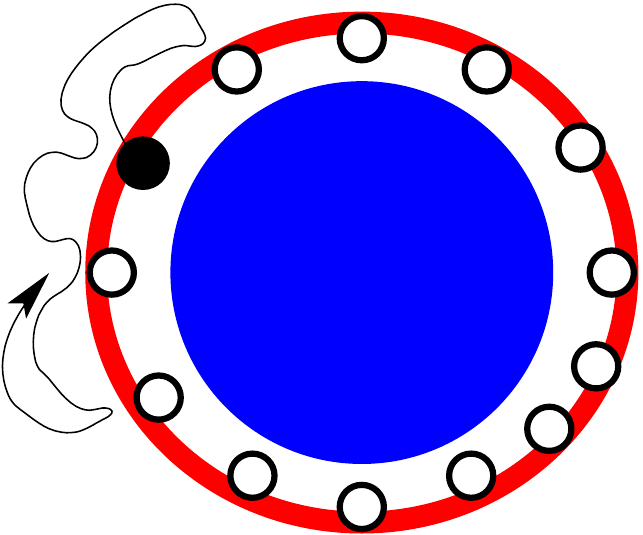}
\caption{
Schematic illustration of the model system.
A ring made up of $N$ sites is immersed in a ``cold" bath 
(represented by inner blue circle) 
and subjected to a ``hot" driving source
(represented by an outer red circle).
The latter has an intensity~$\nu$ 
that can be easily controlled experimentally. 
\rmrk{The transitions rates between the sites 
of the ring are given by \Eq{e1}.  
The dynamics can be optionally regarded 
as that of a random walker in a random environment.} 
After a transient a NESS is reached with current $I(\nu)$.   
} 
\label{f0}
\end{figure}

\rmrk{In this work} 
we consider a geometrically closed mesoscopic system 
that has a non-trivial topology. The system is immersed in 
a finite temperature ``cold" bath.  
Additionally it is coupled to a driving-source,  
\rmrk{with couplings that are log-wide distributed.} 
The driving-source can be regarded 
as a ``hot bath" of infinite temperature. 
Consequently detailed-balance is spoiled, 
and after a transient a NESS is reached. 
Specifically we consider the simplest possible model: 
a mesoscopic ring that is made up of~$N$ sites. 
See \Fig{f0} for a graphical illustration.
Due to the lack of detailed-balance a circulating current is induced.
We shall see that the value of the current \rmrk{($I$)} depends in a non-linear way 
on the intensity \rmrk{($\nu$)} of the driving source. 
Our interest is in the statistical aspects of this dependence.

%%% remarks

Our model is physically motivated and significantly differs 
from the standard setup that has been assumed in past literature. 
Previous study of Sinai-type disordered systems \cite{sinai2}, 
has considered an open geometry with uncorrelated transition rates 
that have the same coupling everywhere. Consequentially 
the random-resistor-network aspect (which is related to local 
variation of the couplings) has not emerged.
Furthermore, in the physically motivated setup that we have 
defined above (ring+bath+driving) Sinai physics would not arise 
if the couplings to the driving source were merely disorderly random. 
The log-wide distribution is a crucial ingredient. 
Finally, in a closed (ring) geometry, unlike an open (two terminal) geometry, 
the statistics of~$I$ is not only affected by the distribution of transition rates, 
but also by the spatial profile of the NESS. 
This is like ``canonical" as opposed to ``grand canonical" setting, 
leading to remarkably different results.

%%%%%%%%%%%%%%%%%%%%%%%%%%%

%%%%%%%%%%%%%%%%%%%%%%%%%%%%%%%%%%%
\rmrk{{\em Outline.-- } 
In Sec.~\ref{sec:model} we describe our minimal model:
a ring coupled to a heat bath and to a driving field, 
with log-wide distribution of coupling.
In Sec. \ref{sec:signchange} we estimate the number of sign changes 
of the steady state current $I(\nu)$ as the intensity of the driving 
is increased.
In Secs. \ref{sec:bonds} and \ref{sec:formula} we present 
an explicit formula for the NESS. This formula is employed 
in Sec. \ref{sec:stat} to study the  statistical properties 
of $I(\nu)$ for an ensemble of rings. 
Specifically, the statistics outside of the Sinai regime 
is investigated in Sec.\ref{sec:out}, while the statistics 
in the Sinai regime is studied in Sec.\ref{sec:in}. 
In the latter case we show how the fingerprints of Sinai physics 
can be extracted from the analysis of $I(\nu)$ curves. 
The results are summarized in Sec. \ref{sec:summary}.}

%%%%%%%%%%%%%%%%%%%%%%%%%%%%%%%%%%%%%%%%%%%%%%%%%%%%%%%%%%%%%%%%%%%%%%%%%%%%%%%%%
%%%%%%%%%%%%%%%%%%%%%%%%%%%%%%%%%%%%%%%%%%%%%%%%%%%%%%%%%%%%%%%%%%%%%%%%%%%%%%%%%
\section{The model}
\label{sec:model}
Consider a ring that consists of sites labeled by~$n$ 
with positions ${x=n}$ that are defined modulo~$N$. 
The bonds are labeled as ${\overrightarrow{n}\equiv(n{-}1 \leadsto n)}$.
The inverse bond is $\overleftarrow{n}$, and if direction does 
not matter we label both by $\bar{n}$. The position of the $n$th bond 
is defined as $x_n \equiv n{-}(1/2)$. The on-site energies $E_n$ 
are normally distributed over a range $\Delta$,  
and the transitions rates are between nearest-neighboring sites:   
\be{1}
w_{\overrightarrow{n}} \ \ = \ \ w^{\beta}_{\overrightarrow{n}} + \nu g_{\bar{n}}
\eeq 
Here $w^{\beta}$ are the rates that are induced by a bath that has 
a finite temperature $T_B$. The $g_{\bar{n}}$ are 
couplings to a driving source that has an intensity~$\nu$. 
These couplings are log-box distributed within ${[g_{\text{min}},g_{\text{max}}]}$.
This means that $\ln(g_{\bar{n}})$ are distributed uniformly 
over a range ${\sigma=\ln(g_{\text{max}}/g_{\text{min}})}$. 
The bath transition rates satisfy detailed-balance, namely 
\beq
\frac{w^{\beta}_{\overrightarrow{n}}}{w^{\beta}_{\overleftarrow{n}}} 
\ \  = \ \ \exp\left[-\frac{E_{n}{-}E_{n{-}1}}{T_B}\right]
\eeq
Assuming ${\Delta \ll T_B}$ one obtains the following approximation: 
\beq
w^{\beta}_{\overrightarrow{n}} \ &\approx& \ \left[1-\frac{1}{2}\left(\frac{E_n-E_{n{-}1}}{T_B}\right)\right]\bar{w}_{\bar{n}}^{\beta} \\ 
w^{\beta}_{\overleftarrow{n}} \ &\approx& \ \left[1+\frac{1}{2}\left(\frac{E_n-E_{n{-}1}}{T_B}\right)\right]\bar{w}_{\bar{n}}^{\beta}
\eeq

The driving spoils the detailed-balance. 
We define the resulted stochastic field as follows:
\be{100} 
\mathcal{E}(x_n) \ \ \equiv \ \ \ln \left[\frac{w_{\overrightarrow{n}}}{w_{\overleftarrow{n}}}\right] 
\eeq
Assuming  ${\Delta \ll T_B}$ we get the following approximation: 
\beq
\frac{w_{\overrightarrow{n}}}{w_{\overleftarrow{n}}} 
\ = \ \frac{w^{\beta}_{\overrightarrow{n}}+\nu g_{\bar{n}}}{w^{\beta}_{\overleftarrow{n}}+\nu g_{\bar{n}}}
\ \approx \ 1- \frac{(E_n-E_{n{-}1})/T_B}{1+(g_{\bar{n}}/\bar{w}_{\bar{n}}^{\beta})\nu}
\eeq
leading to 
\beq
\mathcal{E}(x_n) \ \ \approx \ \ - \left[ \frac{1}{1+g_{\bar{n}}\nu} \right] \frac{E_n{-}E_{n{-}1}}{T_B}
\eeq
In the last equality, without loss of generality,  
the $g_{\bar{n}}$ have been re-scaled such that 
all the bath-induced transitions 
have the same average value ${\bar{w}^{\beta}=1}$.

%%%%%%%%%%%%%%%%%%%%%%%%%%%%%
\begin{figure}

(a)

\includegraphics[width=\hsize]{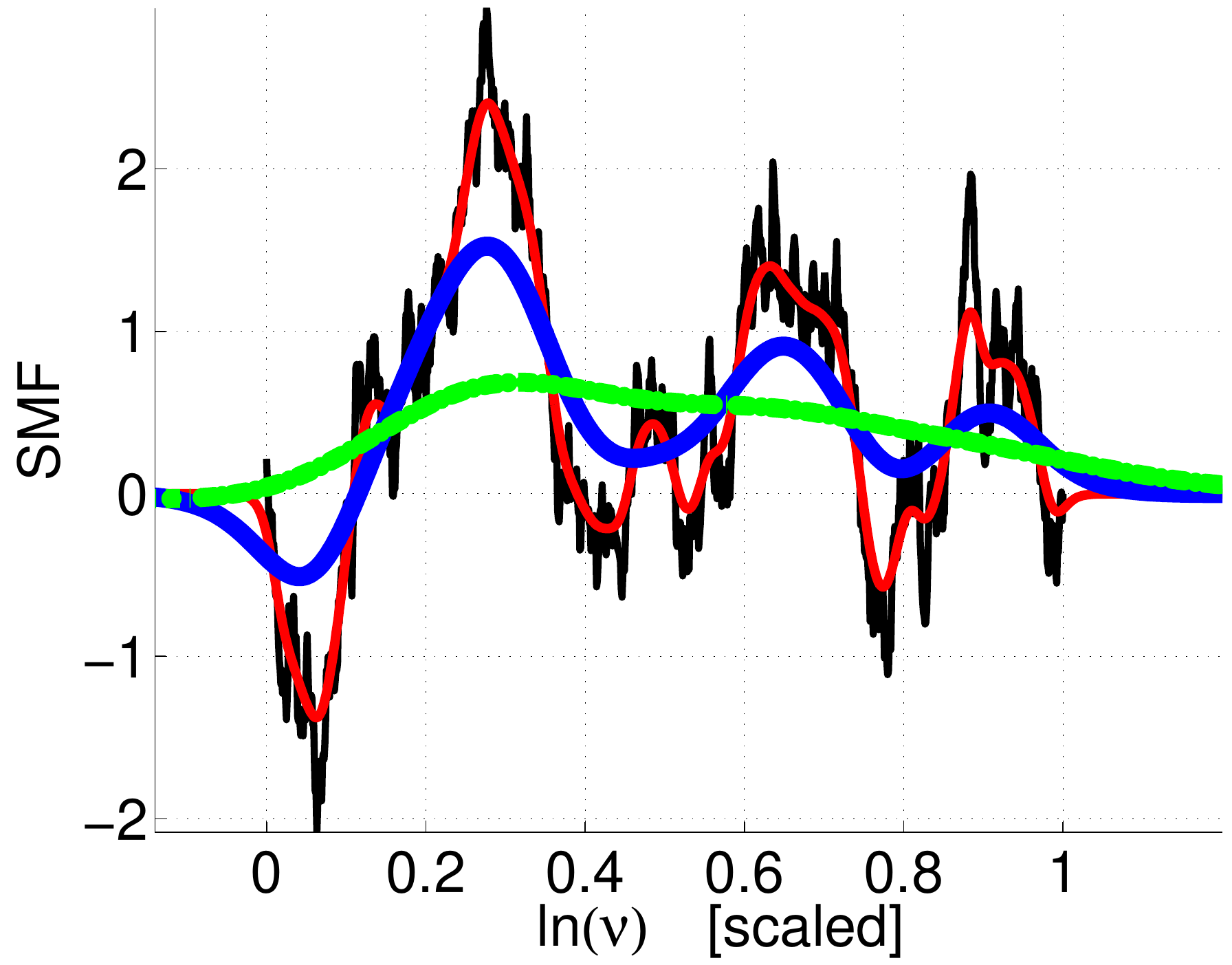}

\vspace*{5mm}

(b)

\includegraphics[width=\hsize]{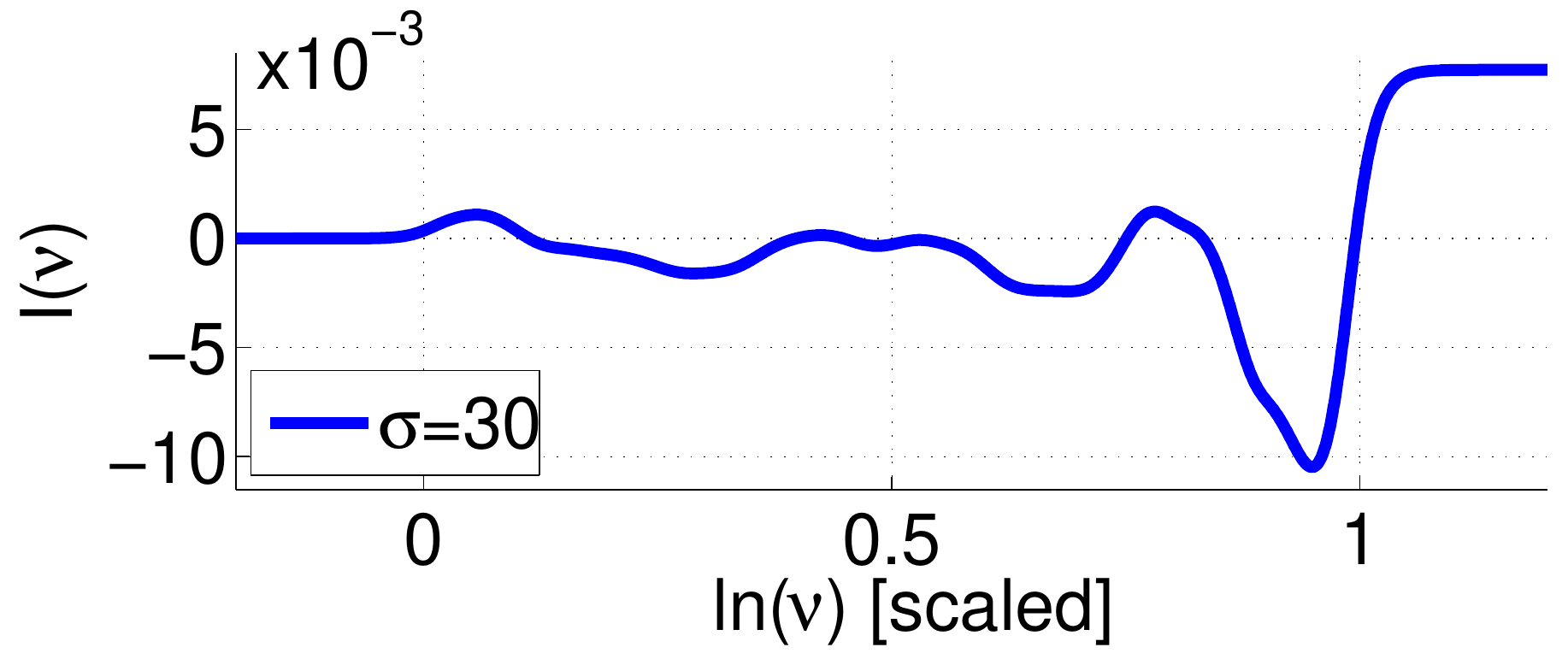}

\vspace*{5mm}

(c)

\includegraphics[width=\hsize]{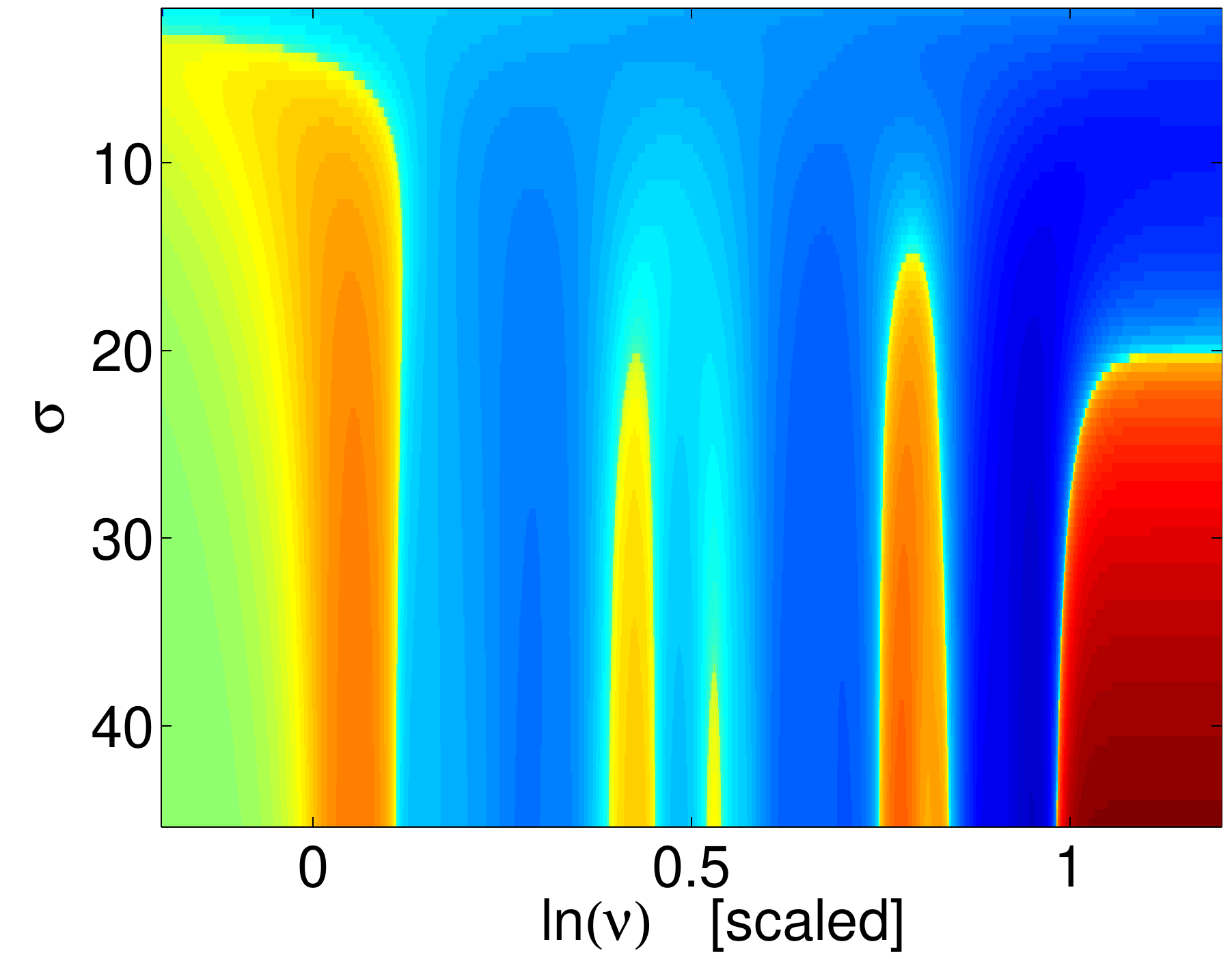}

\caption{
We consider a ring with ${N=1000}$ sites whose energies 
are normally distributed with dispersion ${\Delta=1}$.
The bath temperature is $T_B=10$. 
In (a) the SMF of \Eq{e102} is plotted for $\sigma=\infty$, 
and for $\sigma=50,10,4$. The smaller $\sigma$, 
the smoother $\nu$ dependence.
In (b) a representative $I(\nu)$ curve is plotted.
In (c) a set of $I(\nu)$ curves is color-imaged:
each row is $I(\nu)$ for a different $\sigma$,  
blue and red are for positive and negative (clockwise) 
circulating current respectively.     
In all panels the horizontal axis is 
the scaled driving intensity as defined in \Eq{e1021}.} 
\label{f2}
\end{figure}
%%%%%%%%%%%%%%%%%%%%%%%%%%%%%%

%%%%%%%%%%%%%%%%%%%%%%%%%%%
\section{Current sign reversals in the Sinai regime}
\label{sec:signchange}

The direction of the current $\text{sign}(I)$ is determined 
by the stochastic motive force (SMF), also known as the affinity, 
or as the entropy production \cite{eprd1,eprd2,eprd3,eprd4}:
\beq
\mathcal{E}_{\circlearrowleft} 
\ \ \equiv \ \ \ln \left[\frac{ \prod_n w_{\overrightarrow{n}}}{ \prod_n w_{\overleftarrow{n}}}\right] 
\ \ = \ \ \oint \mathcal{E}(x) \, dx 
\eeq
In the second equality we formally regard $x$ as a continuous variable.
This will make the later mathematics more transparent.  
Assuming  ${\Delta \ll T_B}$ we get the following approximation: 
\be{2001}
\mathcal{E}_{\circlearrowleft} 
\ \ \approx \ \  
-\sum_{n=1}^N \left[ \frac{1}{1+g_{\bar{n}}\nu} \right] \frac{\Delta_n}{T_B}
\eeq
One observes that for ${\nu \ll g_{\text{max}}^{-1}}$ 
the SMF is linear ${\mathcal{E}_{\circlearrowleft}\propto \nu}$,  
while for ${\nu \gg g_{\text{min}}^{-1}}$ it vanishes ${\mathcal{E}_{\circlearrowleft}\propto 1/\nu}$.
In the intermediate regime, which we call below {\em the Sinai regime},  
the SMF changes sign several times, see \Fig{f2}. 
Using the notations
\be{1021}
\tau \ \ \equiv \ \ \frac{1}{\sigma}\ln(g_{\text{max}}\nu)
\eeq
and $\tau_n = (1/\sigma)\ln(g_{\text{max}}/g_{\bar{n}})$,  
the expression for the SMF takes the following form:
\be{102}
\mathcal{E}_{\circlearrowleft}(\tau) = -\sum_{n=1}^N f_{\sigma}(\tau-\tau_n) \frac{E_n{-}E_{n{-}1}}{T_B}
\eeq 
where $f_{\sigma}(t)\equiv [1+\eexp{\sigma t}]^{-1}$ 
\rmrk{drops monotonically from unity to zero like a smoothed step function}.
If $f(t)$ were a sharp step function it would follow
that in the Sinai regime $\mathcal{E}_{\circlearrowleft}(\tau)$ 
is formally like a random walk \cite{rw1,feller,dwass}. 
The number of sign reversals equals the number of times 
the random walker crosses the origin. We have here a coarse-grained 
random walk: the $\tau_n$ are distributed uniformly over a range ${[0,1]}$,  
and each step is smoothed by $f_{\sigma}(t)$ such that the 
effective number of coarse-grained steps is $\sigma$.  
Hence we expect the number of sign changes to be not $\sim\sqrt{\pi N}$ 
but $\sim\sqrt{\pi \sigma}$, reflecting the log-width of the distribution.

%%%%%%%%%%%%%%%%%%%%%%%%%%%%%%%%%%%%%%%%%%
\section{Adding bonds in series}

\label{sec:bonds}

The NESS equations are quite simple and can be solved using elementary 
algebra as in \cite{Wichmann,Roling1,Nitzan,ner}, 
or optionally using the network formalism for stochastic systems \cite{net1,net2,net3}.
Below we propose a generalized resistor-network approach 
that allows to obtain a more illuminating version for the NESS, 
that will provide  better insight for the statistical analysis.  
Let us assume that we have a NESS with a current $I$. 
The steady state equations for two adjacent bonds are 
\beq
I &=& w_{\overrightarrow{1}} p_{0} - w_{\overleftarrow{1}}p_{1} \\
I &=& w_{\overrightarrow{2}} p_{1} - w_{\overleftarrow{2}}p_{2}
\ee
We can combine them into one equation: 
\be{8}
I = \overrightarrow{G} p_{0} - \overleftarrow{G} p_{2},  
\eeq
with 
\beq 
\overrightarrow{G} &\equiv&  \left[ \frac{1}{w_{\overrightarrow{1}}} + 
\frac{1}{w_{\overrightarrow{2}}} 
\left(\frac{w_{\overleftarrow{1}}}{w_{\overrightarrow{1}}}\right) \right]^{-1}
\\
\overleftarrow{G} &\equiv& 
\left[ \frac{1}{w_{\overleftarrow{2}}} + 
\frac{1}{w_{\overleftarrow{1}}} 
\left(\frac{w_{\overrightarrow{2}}}{w_{\overleftarrow{2}}}\right) \right]^{-1}
\eeq
We can repeat this procedure iteratively.  
If we have $N$ bonds in series we get
\be{9}
\overrightarrow{G} \  \ &=& \ \ \left[ \sum_{m=1}^N \frac{1}{w_{\overrightarrow{m}}} 
\,\exp\left(-\int_0^{m{-}1}\!\!\!\!\!\!\!\mathcal{E}(x)dx \right) \right]^{-1} 
\\ \label{e10}
\overleftarrow{G} \  \ &=& \ \ \left[ \sum_{m=1}^N \frac{1}{w_{\overleftarrow{m}}} 
\,\exp\left(\int_m^{N}\!\!\mathcal{E}(x)dx \right) \right]^{-1} 
\eeq

Coming back to the ring, we can cut it at an arbitrary site~$n$, 
and calculate the associated $G$s. 
It follows that ${I=(\overrightarrow{G}_n- \overleftarrow{G}_n) \, p_{n}}$.
Consequently the NESS is
\beq
p_{n} \ \ = \ \ \frac{I}{\overrightarrow{G}_n- \overleftarrow{G}_n} 
\eeq
and  $I$ can be regarded as the normalization factor:
\be{11}
I \ = \ \left[\sum_{n=1}^N \frac{1}{\overrightarrow{G}_n-\overleftarrow{G}_n}\right]^{-1}
\eeq    
In the next paragraph we show how to write these results 
in an explicit way that illuminates the relevant physics.

%%%%%%%%%%%%%%%%%%%%%%%%%%%%%%
\begin{figure}
\includegraphics[width=\hsize]{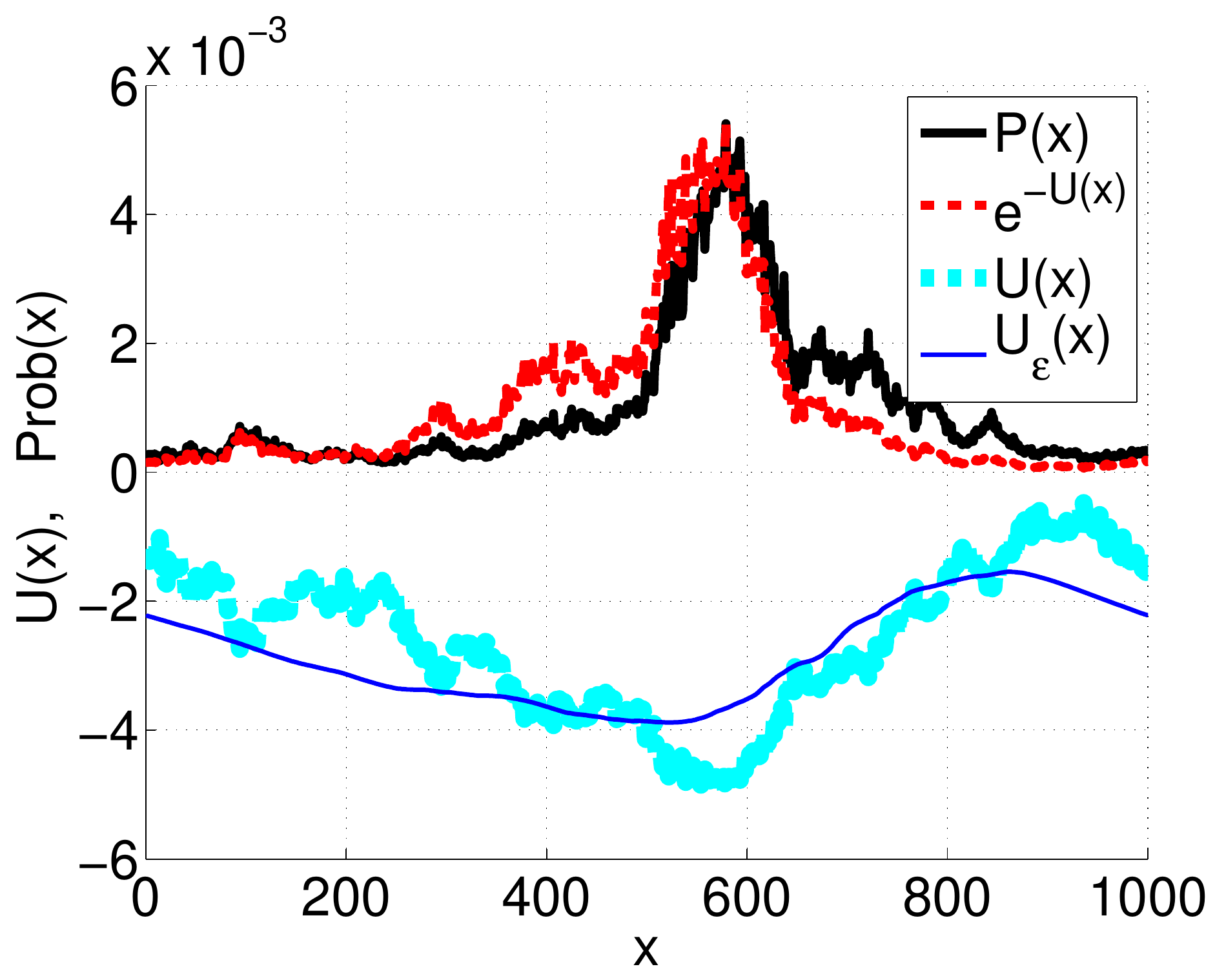}

\caption{
The NESS profile of \Eq{e17} (solid black) 
is similar but not identical to the quasi-equilibrium 
distribution (dashed red line). 
Also shown (lower curves) is the potential landscape $U(x)$ 
and its smoothed version $U_{\varepsilon}(x)$. 
The parameters are the same as in \Fig{f2}, 
with $\sigma=10$, 
and driving intensity that corresponds to $\tau=0.3$.
The bonds were re-arranged to have a larger SMF,
namely $\mathcal{E}_{\circlearrowleft}=7.4$.}

\label{f3}
\end{figure}
%%%%%%%%%%%%%%%%%%%%%%%%%%%%%%

%%%%%%%%%%%%%%%%%%%%%%%%%%%%%%%%%%%%%%%%%%
\section{The NESS formula}
\label{sec:formula}

\rmrk{One should notice that \Eq{e9} and \Eq{e10} cannot 
be treated on equal footing due to a miss-match 
between $m$ and $m{-}1$. For this reason 
we introduced an improved convention for the description 
of the bonds.}  
We define the conductance of a bond as the geometric mean 
of the clockwise and anticlockwise transition rates: 
\beq  
w(x_n) \ \ = \ \ \sqrt{ w_{\overrightarrow{n}} w_{\overleftarrow{n}} }
\eeq
Hence $w_{\overrightarrow{n}} = w(x_n) \exp[(1/2)\mathcal{E}(x_n)]$.
Accordingly  \Eq{e9} and \Eq{e10} \rmrk{can be unified and written as} 
\be{13}
\overrightarrow{G}_n = \left[ \sum_{m=n+1}^{N+n} \frac{1}{w(x_m)} 
\,\exp\left(-\int_{n}^{x_m} \!\!\!\mathcal{E}(x)dx \right) \right]^{-1} 
\eeq
Where the implicit understanding is that the summation and the integration 
are anticlockwise modulo $N$. With the new notations it is easy to see 
that ${\overleftarrow{G}_n = \exp(-\mathcal{E}_{\circlearrowleft}) \, \overrightarrow{G}_n}$.
We use the notation $G_n$ for the geometric mean. Consequently 
the formula for the current takes the form 
\beq
I \ \ = \ \ \left[\sum_{n=1}^N \frac{1}{G_n}\right]^{-1} 2\sinh\left(\frac{\mathcal{E}_{\circlearrowleft}}{2}\right)
\eeq 
while $p_n\propto 1/G_n$. Our next task is to find a tractable
expression for the latter. Regarding $x$ as an extended coordinate, 
the potential $V(x)$ that is associated with the field $\mathcal{E}(x)$ 
is a tilted periodic potential. Adding $[\mathcal{E}_{\circlearrowleft}/N] x$
we get a periodic potential $U(x)$, see \Fig{f3}. Accordingly 
\beq
\int_{x'}^{x''} \!\!\!\mathcal{E}(x)dx \ = \ U(x'){-}U(x'') + \frac{\mathcal{E}_{\circlearrowleft}}{N}(x''{-}x')
\eeq  
With any function $A(x)$ we can associate a smoothed version 
using the following definition  
\beq
\sum_{r=1}^N A(x{+}r) \, \eexp{U(x{+}r)- (1/N)\mathcal{E}_{\circlearrowleft}r} \ \equiv \ A_{\varepsilon}(x) \, \eexp{U_{\varepsilon}(x)} 
\eeq
In particular the smoothed potential $U_{\varepsilon}(x)$ is defined by this expression with ${A=1}$. 
Note that without loss of generality it is convenient to have 
in mind ${\mathcal{E}_{\circlearrowleft}>0}$. (One can always flip the $x$~direction).  
Note also that the smoothing scale $N/\mathcal{E}_{\circlearrowleft}$ becomes larger for smaller SMF.
With the above definitions we can write the NESS expression as follows:
\be{17}
p_n \ \propto \ \left( \frac{1}{w(x_n)} \right)_{\varepsilon} \eexp{-(U(n)-U_{\varepsilon}(n))}
\eeq
This expression is physically illuminating, see \Fig{f3}. 
In the limit of zero SMF it coincides, as expected, 
with the canonical (Boltzmann) result. 
For finite SMF the smoothed pre-factor and the smoothed potential
are not merely constants. Accordingly the pre-exponential factor
becomes important and the ``slow" modulation by the Boltzmann factor 
is flattened. If we take the formal limit of infinite SMF 
the Boltzmann factor disappears and we are left with ${p_n \propto 1/w_n}$    
as expected from the continuity equation for a resistor-network.

%%%%%%%%%%%%%%%%%%%%%%%%%%%%%%%%%%%%%%%%%%
\section{Statistics of the current}
\label{sec:stat}

From the preceding analysis it should become clear that 
the formula for the current can be written schematically as 
\be{112}
I(\nu) \ \ \sim \ \  \frac{1}{N} \, w_{\varepsilon} \, \eexp{-B} \, 2\sinh\left(\frac{\mathcal{E}_{\circlearrowleft}}{2}\right)
\eeq
In the absence of a potential landscape ($U(x)=0$) the formula becomes equivalent to Ohm law: 
it is a trivial exercise to derive it if all anticlockwise and clockwise rates are equal 
to the same values $\overrightarrow{w}$ and $\overleftarrow{w}$ respectively, 
hence $w_{\varepsilon}=(\overrightarrow{w} \overleftarrow{w})^{1/2}$, 
and ${\mathcal{E}_{\circlearrowleft}=N\ln(\overrightarrow{w}/\overleftarrow{w})}$.   
In the presence of a potential landscape we have an activation barrier.
Assuming that the current is dominated by the highest peak 
a reasonable estimate would be
\be{113}
B \ \ &=&  \ \ \text{max}\left\{ U(x){-}U_{\varepsilon}(x) \right\} 
\\ \label{e114} 
&\approx& \frac{1}{2} \Big[ \text{max}\{U\} - \text{min}\{U\} \Big]  
\eeq 
The implication of \Eq{e112} with \Eq{e113} for the {\em statistics} of the current 
is as follows: in the Sinai regime we expect that it will reflect 
the {\em log-wide} distribution of the activation factor, 
while outside of the Sinai regime we expect it to reflect the {\em normal} distributions 
of the total resistance~$w_{\varepsilon}^{-1}$, and of the SMF.  

\rmrk{In the following sections we provide a detailed analysis for the statistics of~$I(\nu)$. 
We shall see that contrary to first impression the extraction 
of the fingerprints of the log-normal statistics in the Sinai regime 
requires extra treatment. The bare statistics is in fact normal in all regimes. }

%%%%%%%%%%%%%%%%%%%%%%%%%%%%%%%%%%%%%%%%
\begin{figure}

\centering

\includegraphics[width=0.9\hsize]{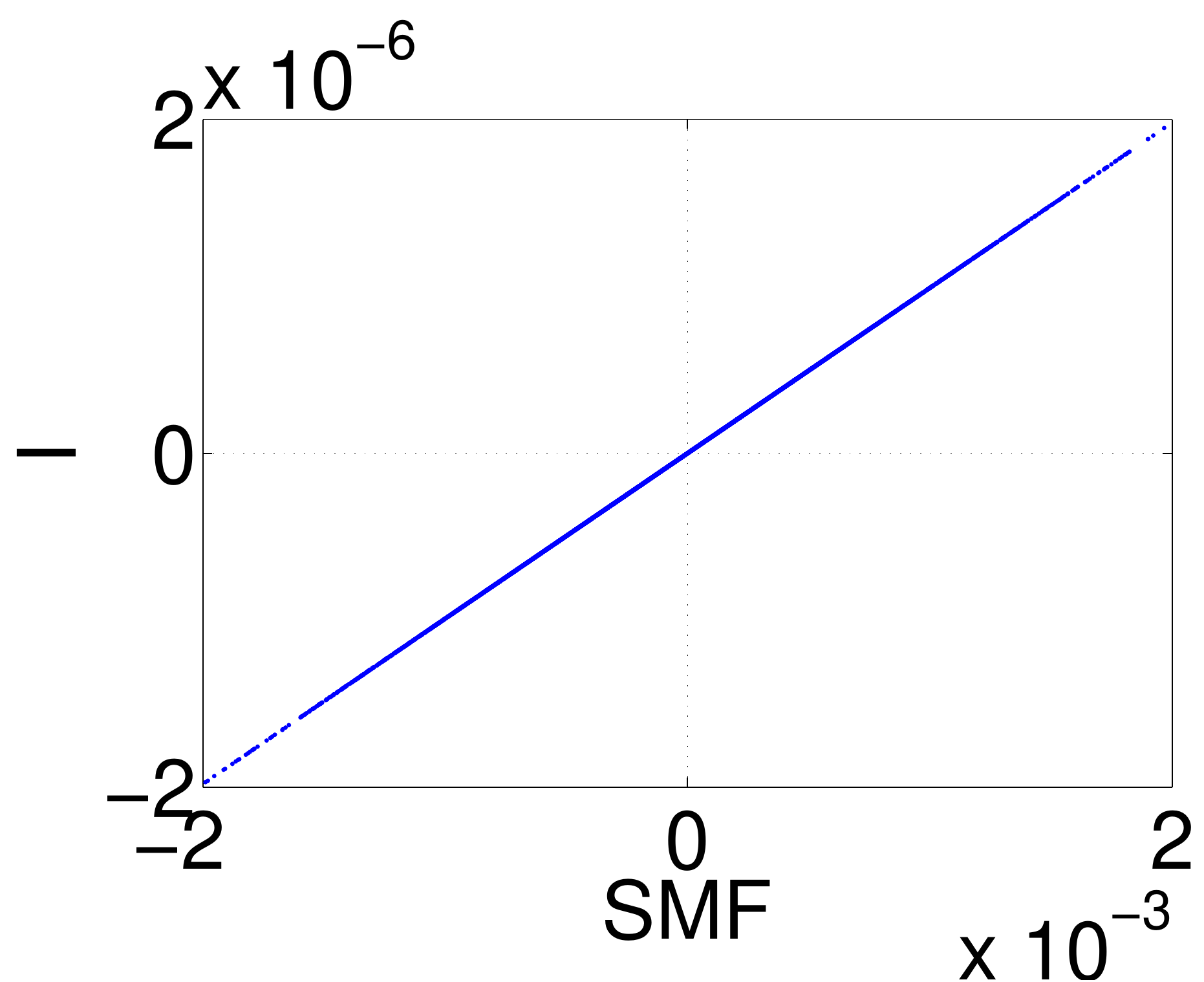}

\includegraphics[width=0.8\hsize]{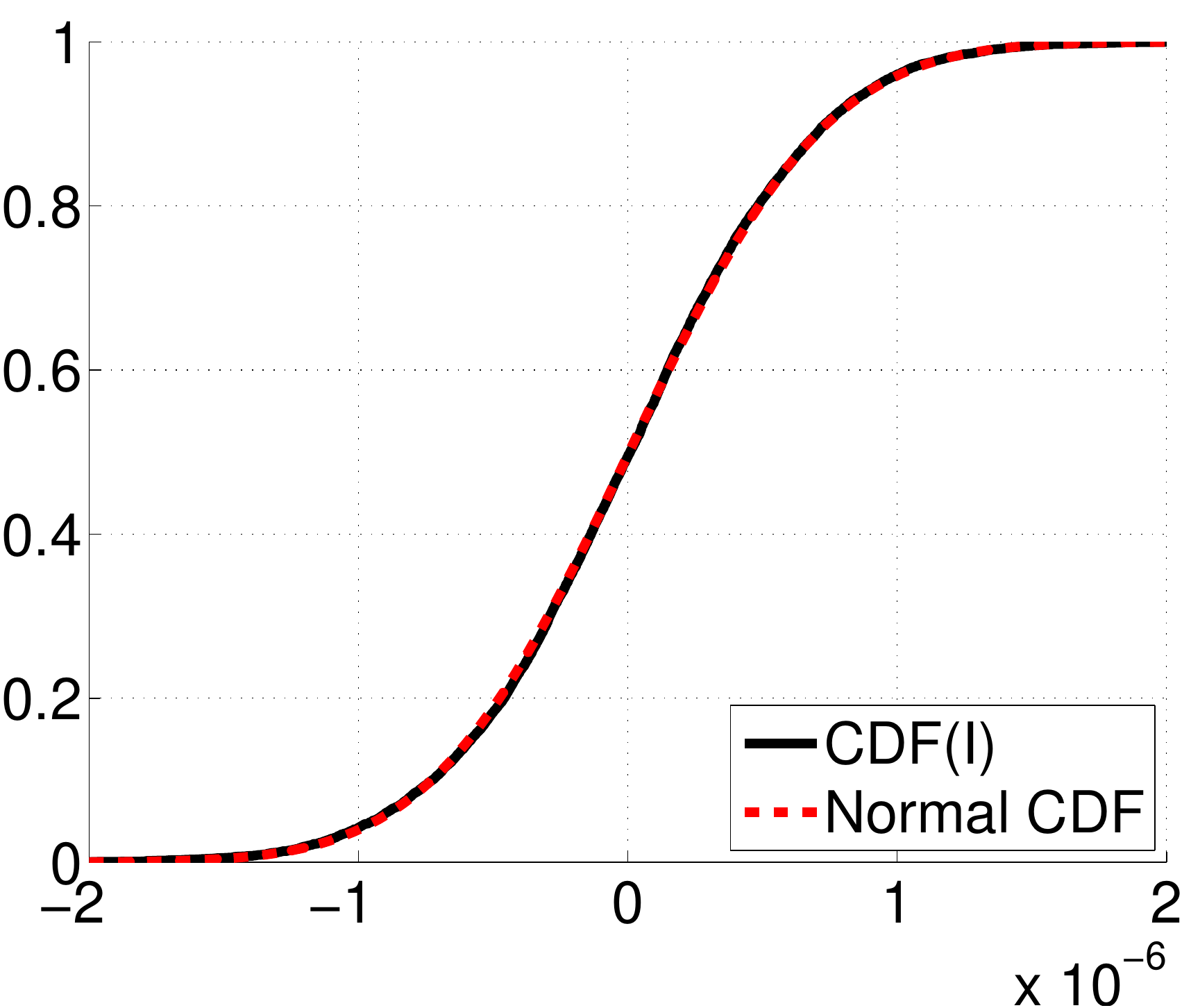}

\caption{
In the linear regime, the current is strongly 
correlated with the SMF (uppper panel), and 
consequently it has {\em normal} statistics (lower panel). 
For the statistical analysis we have generated $10^5$ 
realizations of the ring with $\sigma=6$. 
}
\label{fa1}
\end{figure}
%%%%%%%%%%%%%%%%%%%%%%%%%%%%%%%%%

%%%%%%%%%%%%%%%%%%%%%%%%%%%%%%%%%%%%%%%%%%%%%%%%%%%%%%%%%%%%%%%%%
\section{Statistics of current outside of the Sinai regime}
\label{sec:out}

As the driving intensity is increased one observes a crossover from 
a linear regime, to a Sinai regime, and finally a saturation regime: 
\beq
\text{Linear regime:}     & \ \ \ \ \ & \ \nu < g_{max}^{-1} \\
\text{Sinai regime:}      &&  \ g_{max}^{-1} < \nu < g_{min}^{-1}\\
\text{Saturation regime:} && \ \nu > g_{min}^{-1}
\eeq
Consequently we get for the SMF the following approximations: 
\be{131}
\mathcal{E}_{\circlearrowleft} \ \ \approx \ \ 
\frac{1}{T_B}
\left\{
\amatrix{
\Delta^{(0)}\nu, & \ \ \text{Linear regime} \cr 
-\Delta^{(\infty)}/\nu , & \ \ \text{Saturation regime}
}\right.
\eeq
where 
\beq
\Delta^{(0)} &\equiv&  \sum_{n} g_{\bar{n}}  \Delta_n 
\ \ \sim  \ \ \pm \Big[2N \, \mbox{Var}(g)\Big]^{1/2} \Delta
\\
\Delta^{(\infty)} &\equiv& \sum_{n} \frac{1}{g_{\bar{n}}}  \Delta_n
\ \ \sim \ \ \pm \Big[2N \, \mbox{Var}(g^{-1})\Big]^{1/2} \Delta
\eeq
The estimates for $\Delta^{(0)}$ and for $\Delta^{(\infty)}$
follow from the observation that we have sums of independent 
random variables. For example $\Delta^{(0)}$ can be re-arranged 
as ${\sum_{n=1}^{N} (g_{\bar{n}+1}-g_{\bar{n}}) E_n}$.
Furthermore, we conclude that both $\Delta^{(0)} $ and $\Delta^{(\infty)}$
have {\em normal} statistics as implied by the 
central limit theorem. Consequently we expect {\em normal} statistics 
for the SMF, and hence for the current, as verified in \Fig{fa1}.

%%%%%%%%%%%%%%%%%%%%%%%%%%%%%%%%%
\begin{figure}

\hspace*{-10mm} (a) \\
\includegraphics[width=0.8\hsize]{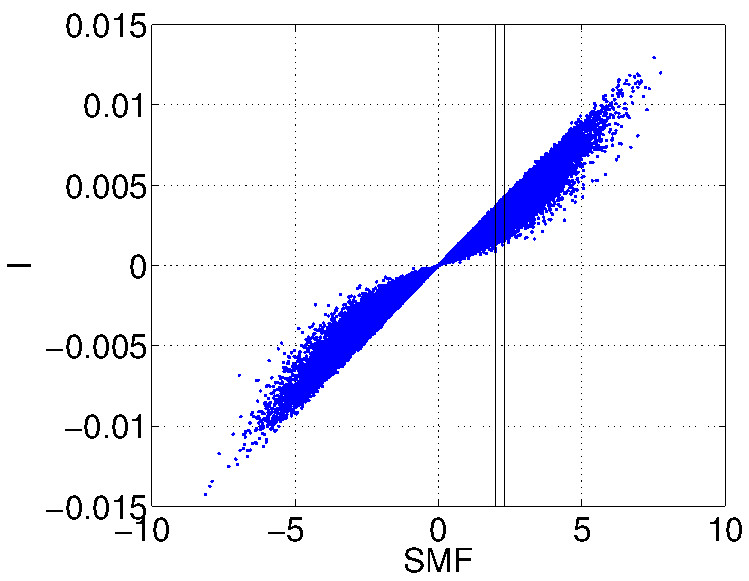}

\ \\

\hspace*{-10mm} (b) \\
\includegraphics[width=0.8\hsize]{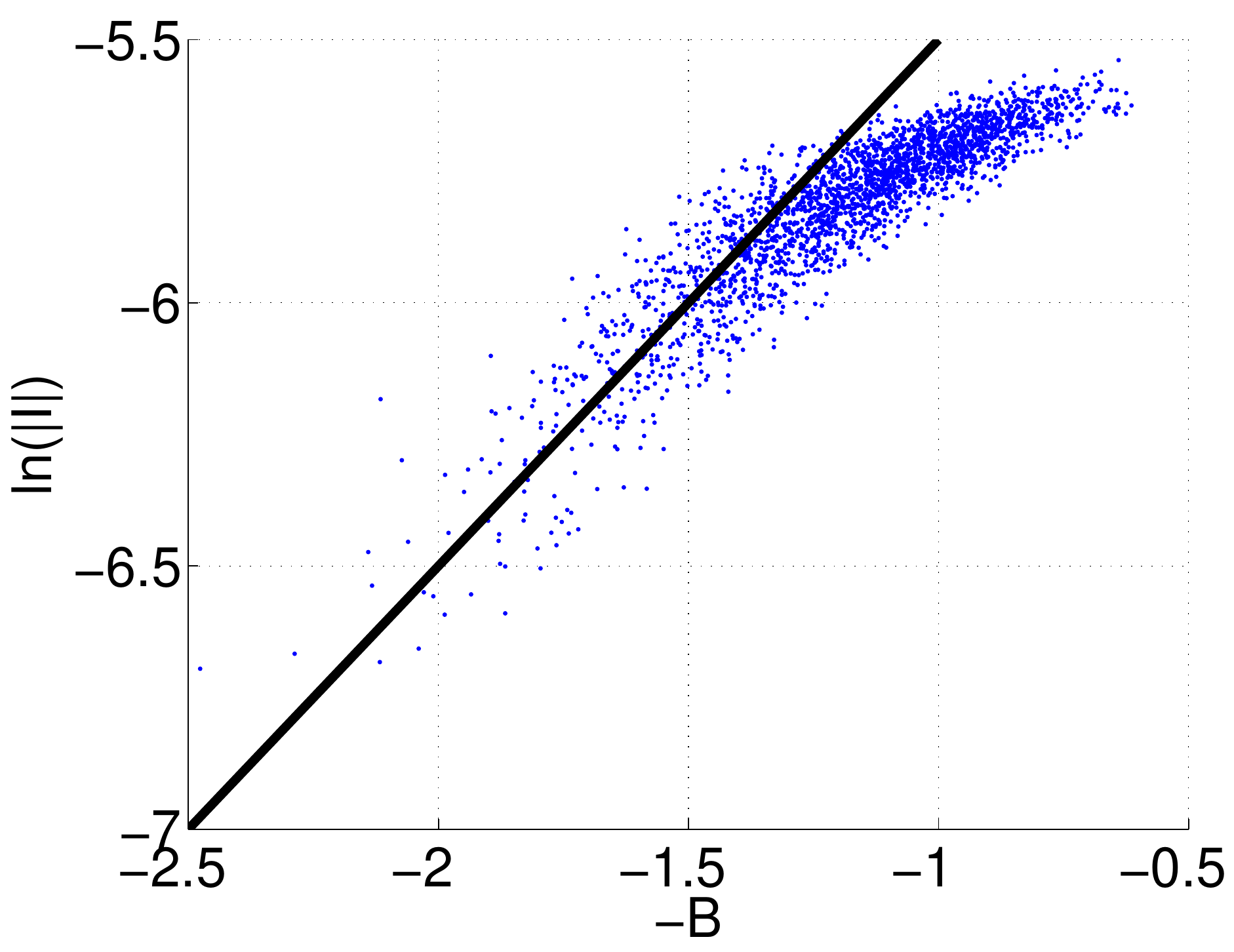}

\caption{(a) Scatter diagram of the current versus the SMF in the Sinai regime.
Note that in the linear regime, see \Fig{fa1}, it looks like a perfect linear 
correlation with {\em negligible} transverse dispersion.  
(b) The correlation between the current~$I$ and the barrier~$B$,  
within the slice ${\mathcal{E}_{\circlearrowleft} \in [2.0,2.1]}$. 
One deduces that the single-barrier approximation is valid for small currents.}

\label{f4}
\end{figure}
%%%%%%%%%%%%%%%%%%%%%%%%%%%%%%%%%

%%%%%%%%%%%%%%%%%%%%%%%%%%%%%%%%%
\begin{figure}
\includegraphics[width=\hsize]{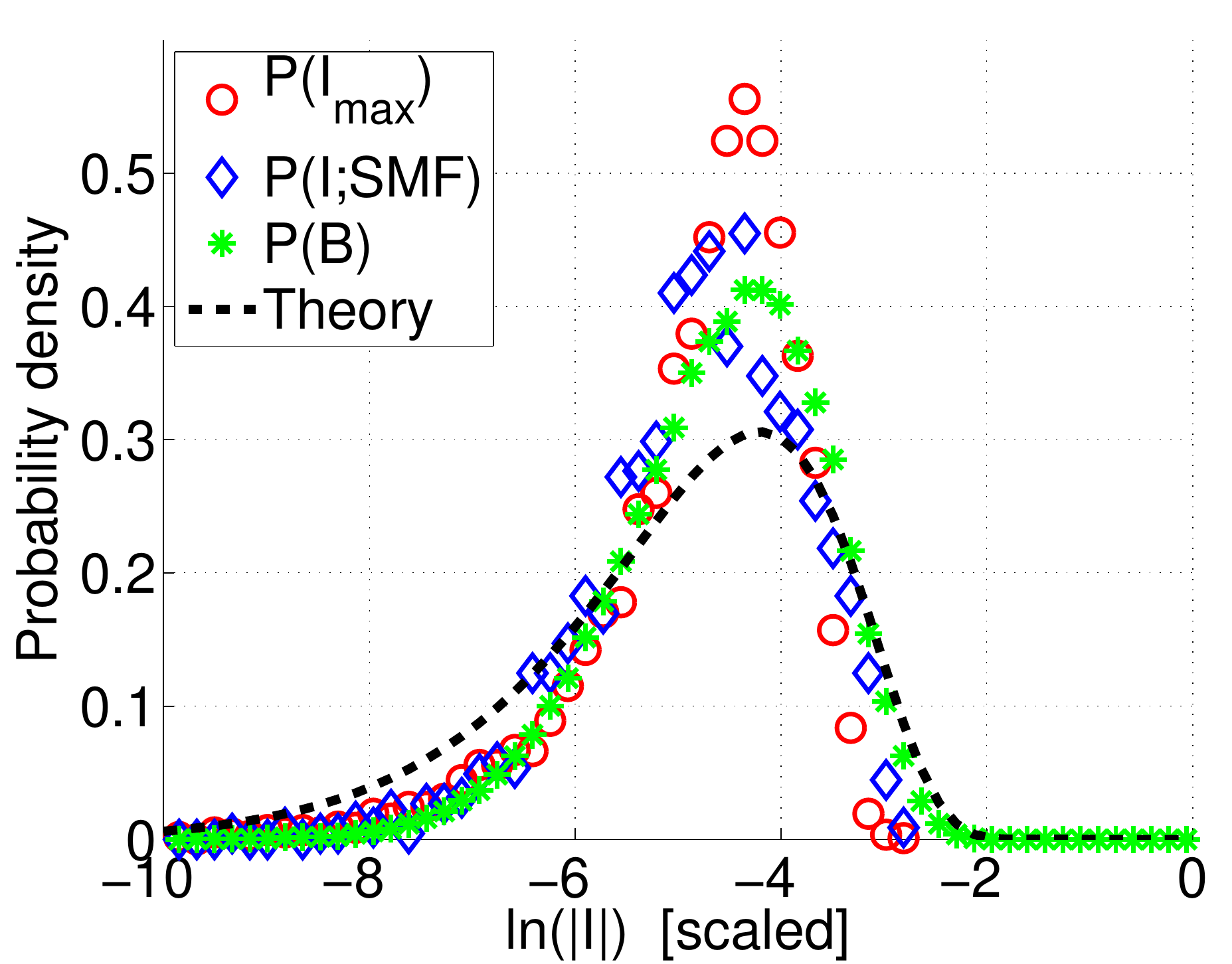}

\caption{The log-wide distribution $P(I)$ of the current 
in the Sinai regime is revealed provided a proper procedure 
is adopted. For theoretical analysis it is convenient 
to plot an histogram of the~$I$ values for a given SMF:
the blue diamonds refer to the data of \Fig{f4}b.  
In an actual experiment it is desired to extract
statistics from $I(\nu)$ measurements without 
referring to the SMF: the red empty circles 
show the statistics of the first maximum of $I(\nu)$.    
Both distributions look the same, and reflect 
the barrier statistics (full green circles). 
The line is the exact version of \Eq{e211}.}

\label{f5}
\end{figure}
%%%%%%%%%%%%%%%%%%%%%%%%%%%%%%%%%

%%%%%%%%%%%%%%%%%%%%%%%%%%%%%%%%%
\begin{figure}

\includegraphics[width=\hsize]{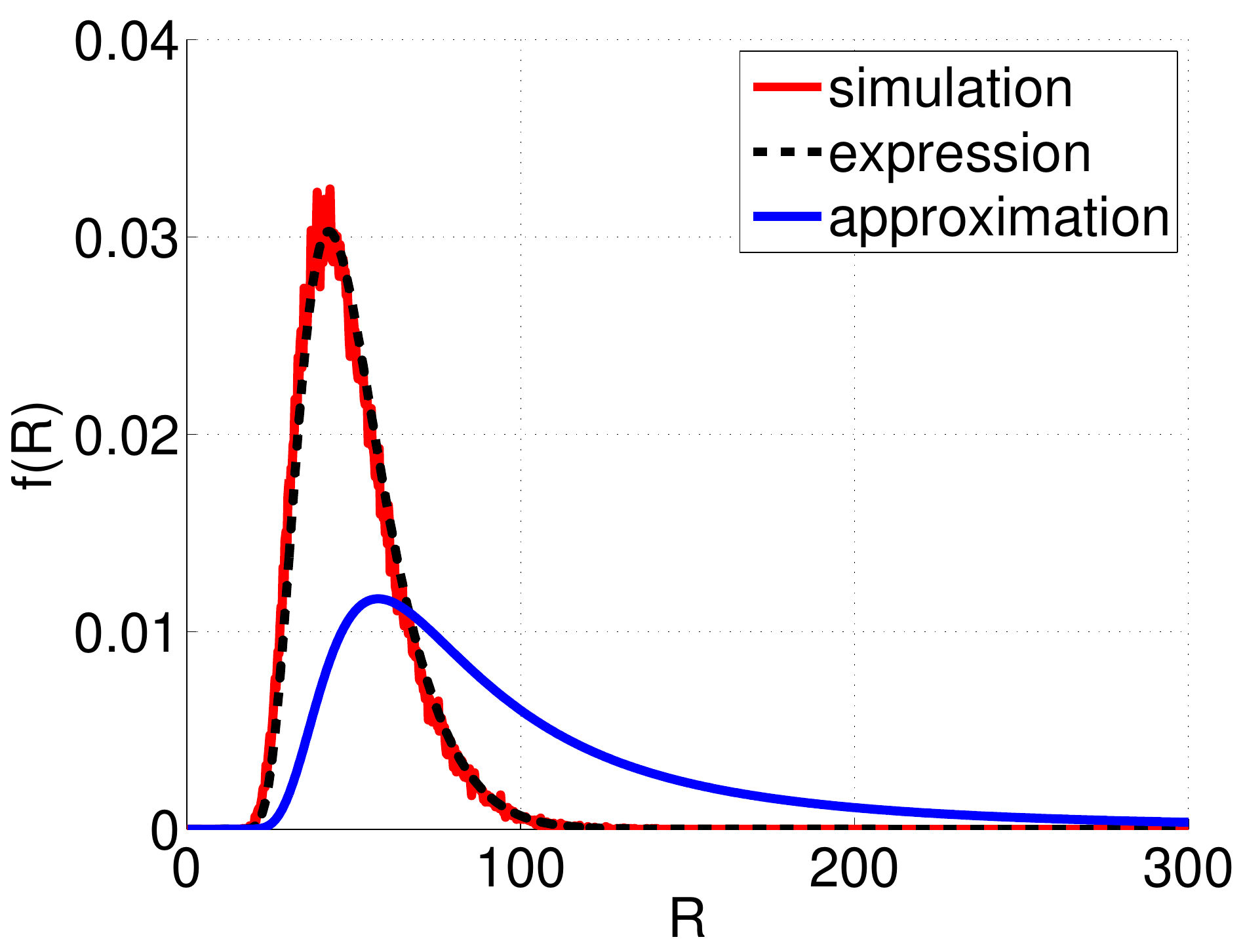}

\caption{
Plot of $f(R)$. Red line is the outcome of a random walk simulation with ${t=1000}$ 
steps that are Gaussian distributed with unit dispersion.   
The black dashed line is the exact result \Eq{e900}, 
while the lower (blue) solid line is from the simple asymptotic approximation \Eq{e937}.} 

\label{fb}
\end{figure}
%%%%%%%%%%%%%%%%%%%%%%%%%%%%%%%%%%%%%

%%%%%%%%%%%%%%%%%%%%%%%%%%%%%%%%%%%%%%%%%%
\section{Statistics in the Sinai regime}
\label{sec:in}

We now focus on the statistics in the Sinai regime. 
In order to unfold the log-wide statistics it is 
not a correct procedure to plot blindly the distribution 
of $\ln(|I|)$. Rather one should look on the joint 
distribution ${(\mathcal{E}_{\circlearrowleft},I)}$. 
See \Fig{f4}a. The non-trivial statistics is clearly apparent.
In order to describe it analytically we use 
the single-barrier estimate of \Eq{e113}, 
which is tested in \Fig{f4}b. We see that it 
over-estimates the current for small $B$ values 
(flat landscape) as expected, but it can be trusted 
for large $B$ where the Sinai physics becomes relevant.

\rmrk{In \Fig{f5} we confirm that the probability 
distribution of the current $P(I;\text{SMF})$, for a given SMF,  
is the same as the barrier ${\exp(-B)}$ statistics. 
We therefore turn to find an explicit expression for the latter.}

The probability to have a random walk trajectory $X_n=U(x_n)$ within $[X_a,X_b]$ 
equals the survival probability in a diffusion process  
that starts as a delta function at ${X=0}$ 
with absorbing boundary conditions at $X_a$ and $X_b$.   
Integrating over all possible positions of the walls 
such that ${X_b-X_a=R}$ is like starting with a 
uniform distribution between the walls. From here 
it is straightforward to deduce what is the probability 
distribution function $f(R)$. 
\rmrk{ The result is displayed in \Fig{fb}.
For the derivation of the exact expression see Appendix~A. 
We note that the occupation-range statistics $f(R)$ is very different 
from that of maximal-distance statistics $f(K)$, see Appendix~B.} 

Turning back to the problem under consideration, \Eq{e114} implies 
that the probability to have a barrier~$B$ is the same as the 
probability that $U(x)$ occupies a range ${R=2B}$. 
Hence it is described by the probability distribution 
function $f(R)$ of \Fig{fb}. The derivation in Appendix~A 
leads to the following practical expression,  
\be{20}
\text{Prob}\left\{\text{barrier}<B\right\} \ \sim \ \exp\left[-\frac{1}{2} \left(\frac{\pi \sigma_U}{2B}\right)^2\right] 
\eeq
where the variance $\sigma_U^2 = 2D N$ is determined 
by the diffusion coefficient $D\propto \Delta^2$ 
\rmrk{that characterizes the potential landscape, 
see for example the illustration in \Fig{f3}.}   
Taking into account that for a given~$\nu$ a fraction of the elements
in \Eq{e102} are effectively zero we get 
\be{211}
\sigma_U^2 \ \ = \ \ 2 \Delta^2 N  \, \frac{\ln(g_{\text{max}}\nu)}{\sigma}
\eeq  
The validity of the exact version of \Eq{e20}, 
which is based \Eq{e900} of Appendix~A, 
has been verified in \Fig{f4}. 
No fitting parameters are required.

In an actual experiment it would be desired to 
extract the statistics from the $I(\nu)$ measurements 
without referring to the SMF. \rmrk{In \Fig{f5} we 
show that the statistics of the first maximum 
of $I(\nu)$ is practically the same as $P(I;\text{SMF})$.
This means that a simple statistical analysis of 
``current versus irradiation" curves is enough in order    
to reveal the fingerprints of Sinai-type physics.}

%%%%%%%%%%%%%%%%%%%%%%%%%%%%%%%%%%%%%%%%%%%%%%%%%%%%%%%%%%%%%%%%%%%%%%%%%%%%%%%%%%%%
%%%%%%%%%%%%%%%%%%%%%%%%%%%%%%%%%%%%%%%%%%%%%%%%%%%%%%%%%%%%%%%%%%%%%%%%%%%%%%%%%%%%
\section{Summary}
\label{sec:summary}

We have introduced a generalized ``random-resistor-network"
approach for the purpose of obtaining the NESS current
due to nonsymmetric transition rates. Specifically our 
interest was focused on the NESS of a ``glassy" mesoscopic system. 
The NESS expression clearly interpolates the canonical (Boltzmann) result 
that applies in equilibrium, with the resistor-network result, 
that applies at infinite temperature. 
Due to the ``glassiness" the current has novel dependence 
on the driving intensity, and it posseses unique statistical properties 
that reflect the Brownian landscape of the stochastic potential.
This statistics is related to Sinai's random walk problem, 
and would not arise if the couplings to the driving source 
were merely disordered.

\rmrk{From the point of view of a practical experiment, 
we have assumed that the most accessible measurements 
would be ``current vs irradiation" curves ($I(\nu)$). 
Namely, experiments in which one changes the external driving 
intensity and observe changes in the resulting NESS. 
The Sinai regime manifests in sign reversals
of the current, whose number is estimated in Sec.~\ref{sec:signchange}.}

\rmrk{By repeating such experiments with an ensemble
of macroscopically equivalent rings one may find imprints 
of the Sinai regime in the statistics of the NESS current.
Our results, depicted in \Fig{f5}, suggest that from $I(\nu)$ measurements alone 
one can extract valuable information regarding 
the Brownian landscape of the stochastic potential;
The functional shape of the distribution provides 
an indication for having Sinai-type physics; 
while from its width one can extract the characteristic 
parameters of the disorder.}

%%%%%%%%%%%%%%%%%%%%%%%%%%%%%%%%%%%%%%%%%%%%%%%%%%%%%%%%%%%%%%%%%%%%%%%%%%%%%%%%%%%%
%\begin{acknowledgements}

{\bf Acknowledgments.-- }
This research was supported by the Israel Science Foundation (grant No.29/11).
We thank Oleg Krichevsky (BGU) for a useful advice. SR is grateful for support from
the Israel Science Foundation (grant 924/11).

%\end{acknowledgements}

%%%%%%%%%%%%%%%%%%%%%%%%%%%%%%%%%%%%%%%%%%%%%%%%%%%%%%%%%%%%%%%%%%%%%%%%%%%%
%%%%%%%%%%%%%%%%%%%%%%%%%%%%%%%%%%%%%%%%%%%%%%%%%%%%%%%%%%%%%%%%%%%%%%%%%%%%

%%%%%%%%%%%%%%%%%%%%%%%%%%%%%%%%%%%%%%%%%%%%%%%%%%%%%%%%%%%%%%%%%%%%%%%%%%%%%%%%%%%%%
%%%%%%%%%%%%%%%%%%%%%%%%%%%%%%%%%%%%%%%%%%%%%%%%%%%%%%%%%%%%%%%%%%%%%%%%%%%%%%%%%%%%%

\clearpage
\onecolumngrid

\appendix

%%%%%%%%%%%%%%%%%%%%%%%%%%%%%%%%%%%%%%
\section{Random-walk occupation-range statistics}

In this section we derived the probability density function $f(R)$ 
to have a random walk process $x(\cdot)$ of $t$ steps that occupies 
a range $R$. This is determined by the probability 
\beq
P_t(x_a,x_b) \ \ \equiv \ \ \text{Prob}\Big(x_a < x(t') < x_b \ \ \mbox{for any $t'\in [0,t]$}\Big)  
\eeq 
Accordingly the joint probability density that 
a random walker would occupy an interval ${[x_a,x_b]}$ is 
\beq
f(x_a,x_b) \ \ = \ \ -\frac{d}{dx_a}\frac{d}{dx_b}P_t(x_a,x_b)
\eeq
It is convenient to use the coordinates
\beq
X &=& \frac{x_a+x_b}{2} \\
R &=& x_b-x_a
\eeq
Consequently the expression for $f(R)$ is  
\beq
f(R) \ &=& \ \int_{-\infty}^{0} \int_{0}^{\infty} dx_a dx_b  
\ f(x_a,x_b) \ \delta\left(R - (x_b-x_a) \right) 
\\
f(R) \ &=& \ - \int_{-R/2}^{R/2} \left(\frac{1}{4}\partial_X^2-\partial_R^2\right) \ P_t(R,X) \ dX
\eeq
Taking into account that $P_t(R,X)$ and its derivatives 
vanish at the endpoints $X=\pm(R/2)$ we get
\be{932}
f(R) \ = \  \int_{-R/2}^{R/2} \partial_R^2 \ P_t(R,X) \ dX 
\ \ = \ \ \partial_R^2 \Big[ R \ P_t(R) \Big] 
\eeq
where $P_t(R)$ is the survival probability of a diffusion process 
that starts with an initial {\em uniform} distribution, instead of 
a random walk that starts as a delta distribution. 
Optionally we can write 
\be{933}
\text{Prob}(\text{range}<R) \ \ = \ \ \partial_R \Big[ R \ P_t(R) \Big]
\eeq
We now turn to find an explicit expression for $P_t(R)$. 
This is done by solving the diffusion equation. 
Using Fourier expansion the solution is
\beq
\rho_t(x) \ = \  \sum_{n=1,3,5,...}^{\infty} 
\exp\left[-D \left(\frac{\pi n}{R}\right)^2t\right] 
\frac{4}{\pi n R}\sin\left(\frac{\pi n}{R}x\right)
\eeq
For simplicity we have shifted above the domain to ${x\in[0,R]}$. 
For the survival probability we get 
\be{935}
P_t(R) \ \ = \ \  
\int_{0}^{R} \ \rho_t(x) \ dx 
\ \ = \ \
\sum_{n=1,3,5,...}^{\infty} 
\frac{8}{\pi^2 n^2 }
\exp\left[-D \left(\frac{\pi n}{R}\right)^2t\right] 
\eeq
Using \Eq{e935} in \Eq{e932} we get
\be{900}
f(R) \ \ = \ \ 
\frac{8\sigma^2}{R^3}
\sum_{n=1,3,5,...}^{\infty} 
\left[ \left(\frac{\pi \sigma n }{R} \right)^2 - 1 \right]
\exp\left[-\frac{1}{2} \left(\frac{\pi \sigma n }{R} \right)^2 \right]
\eeq
This result is in perfect agreement with the numerical simulation of \Fig{fb}. 
Still we would like to have a more compact expression. 
One possibility is to keep only the first term. 
The other possibility is to approximate the summation by an integral: 
\be{936}
\text{Prob}(\text{range}<R) 
\ \ \approx \ \ 
\frac{2}{\pi^2} \frac{\partial}{\partial R} 
\left[ R \int_1^{\infty}  \frac{dx}{x^2} \exp\left(-\frac{\pi^2Dt}{R^2}x^2 \right) \right] 
\ \ = \ \ 
\exp\left(-\frac{\pi^2Dt}{R^2}\right)
\eeq
Either way we get  
\be{937}
\text{Prob}(\text{range}<R) 
\ \ \sim \ \ 
\exp\left(-\frac{1}{2}\left(\frac{\pi\sigma}{R}\right)^2\right)
\eeq
where ${\sigma^2=2Dt}$.
This asymptotic expression is illustrated in \Fig{fb}.
Though it does not work very well, it has the obvious 
advantage of simplicity.

%%%%%%%%%%%%%%%%%%%%%%%%%%%%%%%%%%%%%%
\section{Random-walk maximal-distance statistics}

The occupation-range statistics of the previous section should not be confused 
with the maximal-distance statistics. The maximal distance from the initial point 
is defined as follows:
\beq
K \ \ = \ \ \max[x(t)], \ \ \ \ \ \ \mbox{where $0<t<N$} 
\eeq
Naively, one might think that the probability distribution of $K$ 
is similar to the probability distribution of $R$ that has been discussed 
in the previous section. But this is not true. Furthermore, 
it is also very sensitive to whether the random walk is constrained 
to end up at the origin, ${x(N)=x(0)=0}$.  Without the latter 
constraint $f(K)$ is finite for small $K$, but if the constraint is 
taken into account, it vanishes linearly in this limit. 

\begin{figure}
\includegraphics[width=7cm]{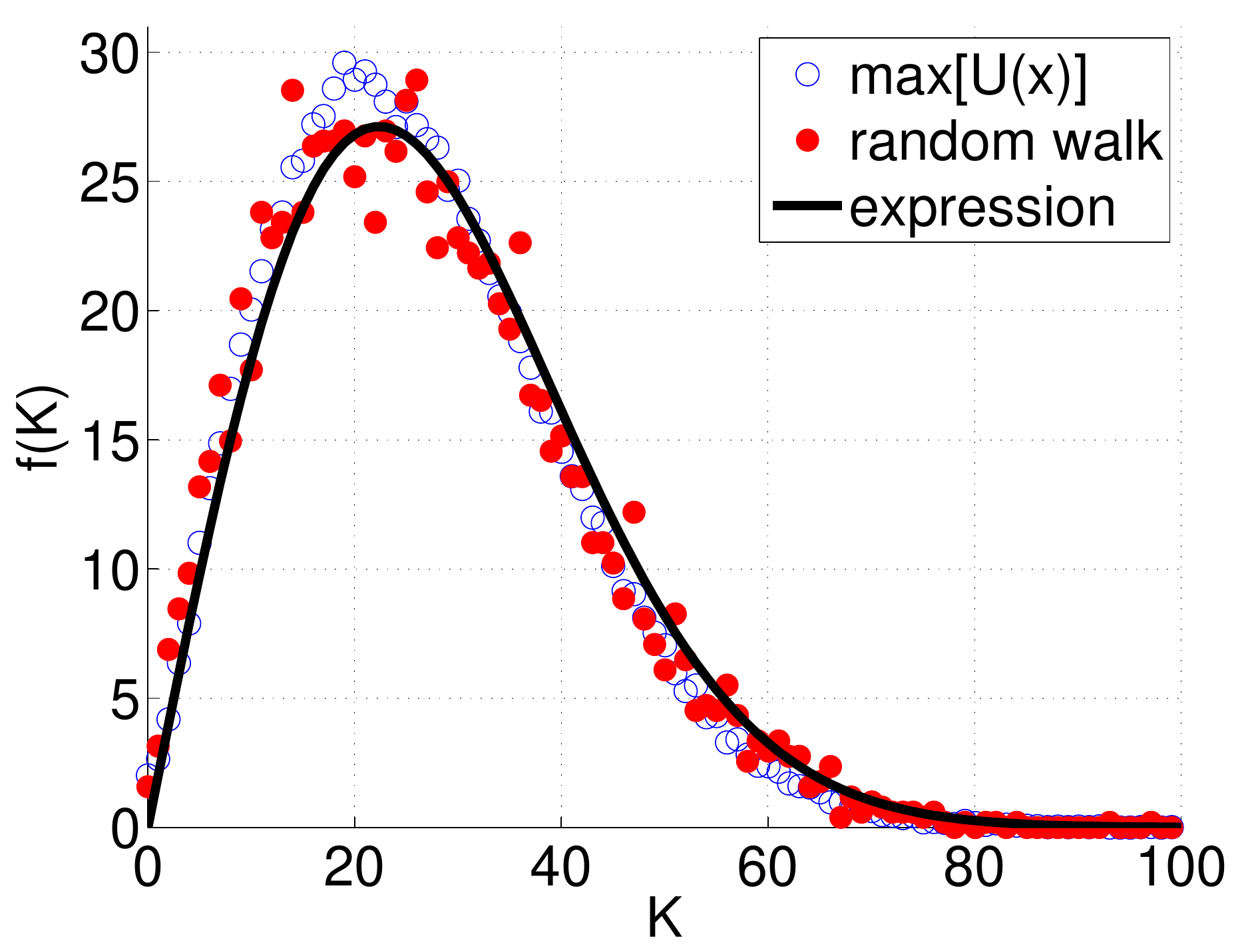} 
\includegraphics[width=7cm]{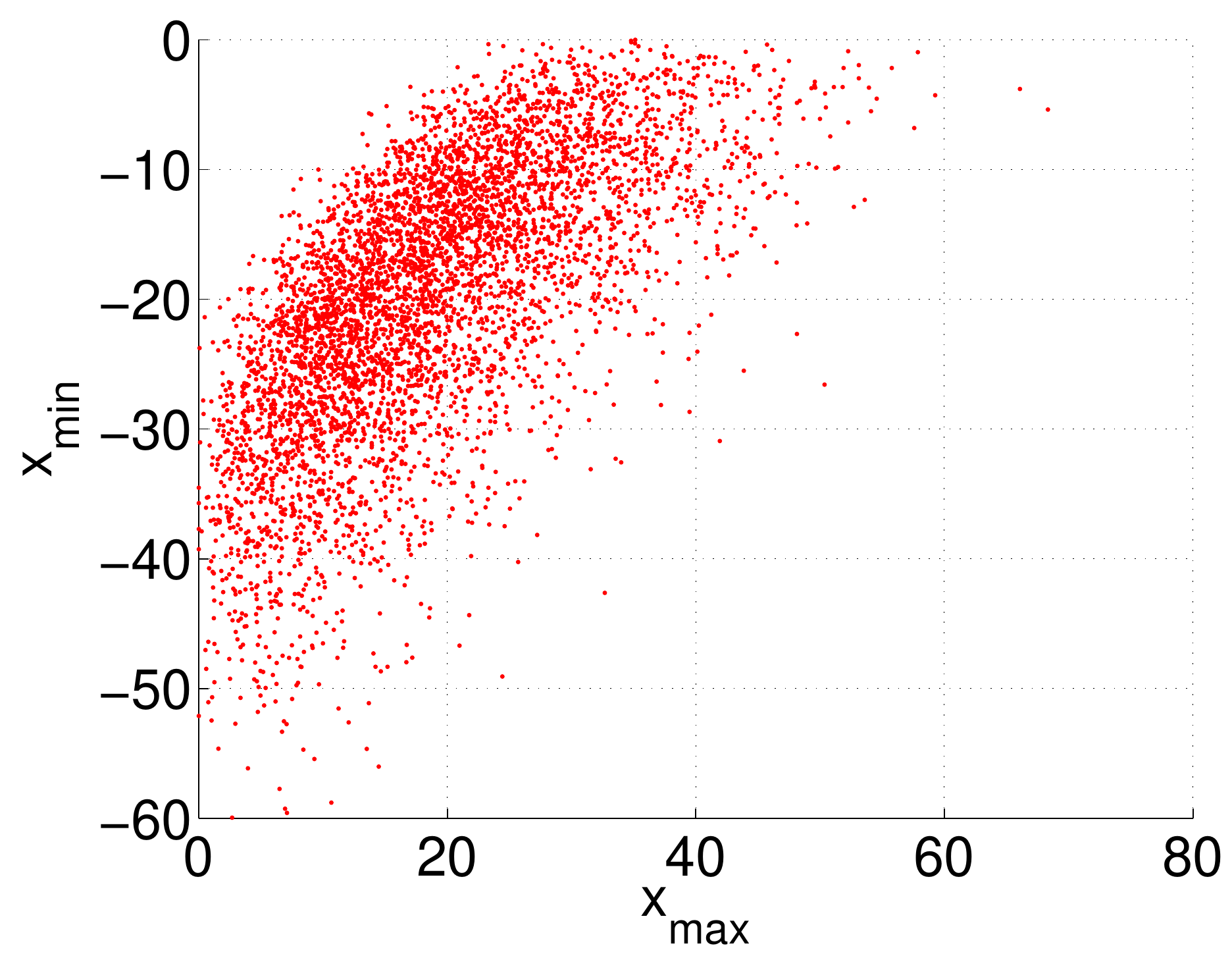} 

\caption{
[Left panel] Plot of $f(K)$. 
The histogram of $\max[U(x)]$ values over many ring realizations (blue circles)  
is compared with the $K$ statistics in a constrained random walk process (red points).
The analytical result \Eq{e51} is represented by a black line.
[Right panel] Scatter plot of ${(x_{\text{min}}, x_{\text{max}})}$ for the 
same random walk simulation illustrating the strong correlation. }

\label{pu}
\end{figure}

It is the constrained random walk process that describes the potential $U(x)$. 
The exact result for the the $K$ statistics in this case is known \cite{dwass}:
\beq
\text{Prob}(K \geq k;N) \ \ = \ \  
\frac
{\left(\begin{array}{c}2N \\N-k\end{array}\right)}
{\left(\begin{array}{c}2N \\N\end{array}\right)},  
\ \ \ \ \ \ \ \ \ k = 0,1,2 \cdots N
\eeq
Switching variables to $\kappa=k/N$ and taking the large $N$ limit, 
one obtains the probability density function 
\be{50}
f(\kappa) = 
N \left[\frac{(1-\kappa)^{\kappa-1}}{(1+\kappa)^{\kappa+1}}\right]^N 
\ln \left[\frac{1+\kappa}{1-\kappa} \right] 
\ee
which has a peak at $\kappa \sim 1/\sqrt{2N}$.
For $\kappa\ll1$ this expression can be approximated 
by the simple function. Switching back to~$K$ 
it takes the form 
\be{51}
f(K) \ \ \approx \ \ \frac{2K}{N} \ \exp\left[-\frac{K^2}{N}\right]
\ee
In \Fig{pu}a we illustrate this distribution and 
demonstrate its applicability to the ${U(x)}$ of the ring model.
In \Fig{pu}b we illustrate the joint distribution of the 
extreme values ${x_{\text{min}}=\min[x(\cdot)]}$ and ${x_{\text{max}}=\max[x(\cdot)]}$.
The $f(R)$ distribution of the previous section corresponds to its projection 
along the diagonal direction, while the $f(K)$ distribution of the present section 
is its projection along the horizontal or vertical directions.

%%%%%%%%%%%%%%%%%%%%%%%%%%%%%%%%
%%%%%%%%%%%%%%%%%%%%%%%%%%%%%%%%%%

\clearpage

\end{document}